\documentclass{aa}
\usepackage{graphicx}
\usepackage{epsfig}
\begin{document}

\newcommand{\kms}  {km s$^{-1}$}
\newcommand{\msol} {M$_{\odot}$}
\newcommand{\teff} {$T_{\rm eff}$}
\newcommand{\vt} {$v_{\rm turb}$}
\newcommand{\sk} {Sk$-$69$^{o}$202} 
\def\lesssim{\mathrel{\hbox{\rlap{\hbox{\lower4pt\hbox{$\sim$}}}\hbox{$<$}}}}
\def\gtrsim{\mathrel{\hbox{\rlap{\hbox{\lower4pt\hbox{$\sim$}}}\hbox{$>$}}}}

 
   \title{The evolutionary status of Sher25 - implications for blue 
          supergiants and the progenitor of SN1987A}

   \author{S. J. Smartt\inst{1} \and D. J. Lennon\inst{2} \and R. P. Kudritzki\inst{3} 
           \and F. Rosales\inst{1} \and R. S. I. Ryans\inst{4} \and N. Wright\inst{1}}

   \offprints{S. J. Smartt: sjs@ast.cam.ac.uk}

   \institute{Institute of Astronomy, University of Cambridge, Madingley Road,
              CB3 OHA, Cambridge, England 
        \and
        The Isaac Newton Group of Telescopes, Apartado de Correos 368, 
        E-38700, Santa Cruz de La Palma, Canary Islands, Spain
        \and
        Institute for Astronomy, University of Hawaii at Manoa,
	2680 Woodlawn Drive, Honolulu, Hawaii 96822 
        \and
	The Department of Pure and Applied Physics, The Queen's University
	of Belfast, Belfast BT7 1NN}
   \date{Received 19-02-02; accepted 03-06-02}

\abstract{
The blue supergiant Sher\,25 in the massive Galactic cluster NGC3603
is surrounded by a striking emission line nebula. The nebula contains
an equatorial ring and probable bi-polar outflows, and is similar in
morphology, mass and kinematics to the shell now visible around
SN1987A. It has been suggested that both nebulae were ejected while
Sher\,25 and the progenitor of SN1987A were in previous red supergiant
phases. In the case of Sher\,25 this is based on the qualitative
strengths of nebular [N\,{\sc ii}] emission which is indicative of
nitrogen enriched gas. This gas may have been dredged up to the
stellar surface by convective mixing during a previous red supergiant
phase. We present optical high-resolution spectra of Sher\,25 and a
model photosphere and unified stellar wind analysis which determines
the atmospheric parameters, mass-loss rate and photospheric abundances
for C, N, O, Mg, and Si. We compare these results, in particular CNO,
to other Galactic B-type supergiants and find that Sher\,25 does not
appear extreme or abnormal in terms of its photospheric nitrogen
abundance. The C/N and N/O ratios are compared to surface abundances
predicted by stellar evolutionary calculations which assume the star
has gone through a red supergiant phase and convective dredge-up. In
particular we find that the N/O abundance is incompatible with the
star having a previous red-supergiant phase, and that the nebulae is
likely to have been ejected while the star was a blue supergiant. The
results are compatible with some degree of rotationally induced mixing
having occurred while the star was on or near the main-sequence. This
is similar to what has recently been found for nebulae surrounding
LBVs.  In addition our wind analysis suggests the star currently has a
relatively normal mass-loss rate in comparison with other Galactic
B-type supergiants and sits comfortably within the wind
momentum-luminosity relationship. In light of the evidence regarding
massive evolved early-type stars in the Galaxy we suggest there is no
object which shows any evidence of having had a previous red
supergiant phase and hence of undergoing blue loops in the HR diagram.
   }
\titlerunning{The blue supergiant Sher25}
\authorrunning{Smartt et al.}

   \maketitle

%

\section{Introduction}

The study of hot luminous B-type supergiant stars is of fundamental 
importance for understanding the late evolutionary stages of massive stars 
and the chemical evolution of our own and external galaxies. Massive stars 
enrich the interstellar medium with the heavy chemical elements during
all stages of their evolution, from fast winds in their early hot 
stages through cool dense outflows from red supergiants and finally 
and most spectacularly during their deaths in  
core-collapse supernovae explosions. The fact that the progenitor of 
SN1987A was identified as a hot B-type supergiant makes them extremely
interesting objects, being the final stage in the life
of at least some very massive stars.

NGC 3603 is a massive H\,{\sc ii} region ionized by a young, luminous
cluster located approximately 7kpc from the Sun in the fourth 
galactic quadrant {({\em
l}=291$^{\circ}$.62, {\em b}=-0$^{\circ}$.52)}. It has been  described as the
most massive optically visible giant H\,{\sc ii} region in the Galaxy
(Goss \& Radhakrishnan \cite{gossradha69}). Given its relative proximity
and the large number of massive stars it contains, this cluster 
provides an opportunity to study the upper mass region of the IMF and 
the evolutionary processes of very massive stars. It contains
approximately 50 O and early B-type stars and the age of the 
original starburst is estimated somewhere between 2-4\,Myr
(Melnick et al. \cite{mtt89}, Moffat \cite{moff83}, 
Crowther \& Dessart \cite{crow98}). The
cluster includes the
trapezium-type system HD97950 which has three WR stars identified and six
O3 stars all within 0.3 pc (Drissen et al. \cite{drissen95}). Of 
particular interest is the star no. 25 reported first by Sher (\cite{sher65}).
Moffat (\cite{moff83}) presented a moderate resolution spectrum of 
this blue supergiant and classified it as a B1.5\,Iab and 
Van den Bergh (\cite{vandb78}) has described it 
as the most highly evolved star in the cluster.

The main reason for the interest in this star is the spectacular
circumstellar nebula, discovered by Brandner et al. (\cite{bran97a})
and beautifully illustrated in HST press release images 
STScI-PRC99-20\footnote{http://oposite.stsci.edu/pubinfo/PR/1999/20/index.html}. It consists of a ring-shaped equatorial emission line nebula (expansion
velocity of 20 \kms) and probable bipolar outflows to the northeast
and to the southwest (expansion velocity of 83 \kms).  The
H$\alpha$+[N\,{\sc ii}] and NIR images of Brandner et
al. (\cite{bran97a}, \cite{bran97b}) indicate that both the ring and
the bi-polar ejecta have roughly the same dynamical age which would
suggest a common origin in a stellar outburst approximately 6600\,yrs
ago. The [N\,{\sc ii}]/H$\alpha$ emission line ratio of the ring
appeared to be higher than that of the background H\,{\sc ii} region,
which Brandner et al. have qualitatively interpreted as evidence for
nitrogen enrichment in the ejecta and hence possibly that the gas has
been processed through the CN and ON-cycles in the core of Sher\,25
during its main-sequence phase. However there are no {\em quantitative}
determinations of electron temperature, density and hence
abundances in either the ring or bi-polar lobes. Sher\,25 is currently
a hot blue supergiant and Brandner et al. have suggested that the star
lived briefly for a period as a red supergiant during which time it
ejected the nebula. In this picture Sher\,25 has undergone a
``blue-loop'' in the HR diagram, first evolving to the red after
main-sequence core-H exhaustion and then back again to the B-type
supergiant region. However these conclusions are based mainly on the
existence and morphology of the nebula and the suggestion that it may
be enriched with nitrogen. How massive stars generally evolve in the
upper regions of the HR diagram is still not well
understood. Theoretical models have many free parameters which can
affect the path of a star after (and in some cases during) the
main-sequence lifetime, such as mass-loss, rotation, convection and
overshooting treatment, and metallicity.  Observational evidence on the
evolutionary status and history of evolved massive stars such as blue
and red supergiants, LBVs and WR stars provide necessary constraints
on the models.  Hence determining the nature of the circumstellar gas
and the evolutionary phase of Sher\,25 is of great interest in this
context. In addition it has been suggested (by Brandner et al.
\cite{bran97b}) that Sher\,25 could be a twin of the progenitor of
SN1987A, which was a B3Ia supergiant (Walborn et
al. \cite{wal89}). This is due to the existence of a similar ring
structure now seen around SN1987A which in existence before the 
SN explosion and hence was ejected by the
progenitor star. Fransson et al. (\cite{fransson89}) suggested that it
had passed through a previous RSG phase given the CNO line strengths
in the early-time UV spectra, although the uncertainties are large. 
Therefore the star Sher\,25 could be a possible 
supernova Type\,II progenitor. 

In the wider picture the existence and relative number of 
blue supergiants in different mass ranges needs to be understood
to provide stronger constraints on evolutionary theory. The 
problem with predicting the correct number ratio of blue-red supergiants
in different metallicity regimes was first pointed out by 
Langer \& Maeder (\cite{lanmed95}). The B/R-ratio appears to 
increase with metallicity and such a trend has not been 
generally reproduced by theory. Recently 
Maeder \& Meynet (\cite{mae2001}) have suggested that introducing 
rotation into the calculations does go some way to improving the 
comparison with observations. If massive stars do go through 
blue-loops then the blue-supergiant population may be made up 
of pre- and post red-supergiant objects (e.g. see models of Schaller
et al. \cite{scha92}).
However distinguishing between two populations and determining
their ratios has been a notoriously difficult problem. One 
method that has been tried is using stellar surface chemistry as an 
indicator of evolutionary status. Qualitatively for example
Jaschek \& Jaschek (\cite{jasjas67}) and later Walborn (\cite{wal72})
suggested that the carbon and nitrogen line strengths in OB stars
could be used to define separate populations, which led to 
the OBN/OBC spectral classification. In this scheme the 
OBN stars generally have stronger nitrogen and weaker carbon 
lines, and vice-versa for the OBC stars. Also there is a wide 
variation in the strength of the differences found between 
stars. These stars are usually explained by invoking some mechanism 
that will mix core processed gas to the surface. This could be 
either dredge up by convective mixing during a previous red 
supergiant phase, or alternatively rotationally induced mixing 
during and just after the main-sequence phase could provide
an efficient mechanism (e.g. Denissenknov 
\cite{den94}, Talon et al. \cite{tal97}, Maeder \& Meynet \cite{mae2001}). 

As the nebulae around Sher\,25 could be suggestive of it 
having definitely gone through a previous red-supergiant
phase, it is of great interest to study its atmospheric and wind
parameters in detail and deduce its surface chemical composition. 
These can then be compared with values estimated from 
different evolutionary scenarios, and also compared to 
the general population of Galactic B-type supergiants. 
A great deal of effort has 
recently been invested in understanding Galactic B-type supergiants
and two papers in particular have set cornerstones in the 
field. The first is the non-LTE model atmosphere
study of atmospheric parameters, line strengths
and abundance trends of McErlean et al. (\cite{mcer99}, hereafter Paper\,I),
which studied a large sample of B-type supergiants in a homogeneous way. 
The difficulties in analysis methods and ways of mitigating errors
were presented which provides an excellent starting point for 
future detailed analysis of single objects or homogeneous groups. 
The second is the extensive study of the wind parameters of 
fourteen early B-type supergiants by Kudritzki et al. (\cite{kud99}, 
hereafter Paper\,II). This used non-LTE unified model atmospheres
to investigate the H$\alpha$ line profiles and 
determine stellar wind properties for a subset of the stars in 
Paper\,I (and some A-type supergiants in addition). A tight 
relationship between the stellar wind momentum and 
luminosity was found, and the Balmer line profiles were 
quantitatively reproduced. 

In this paper we present a non-LTE model photosphere analysis to
deduce the surface parameters and abundances of Sher\,25. In addition, a 
non-LTE unified model atmosphere is used to analyse the 
H$\alpha$ emission line and characterise the stellar wind. 
We place the results of these analysis in the general 
context of the previously analysed large sets of Galactic 
B-type supergiants. 


\section{Observational data}
\label{obsdata}
Spectra of Sher\,25 ($V=12.3, B=13.7$; Brandner et al. \cite{bran97a})
were taken with the UCLES \'{e}chelle spectrometer at the 
Anglo-Australian Telescope on the night of 28th June 1999. The 
79 lmm$^{-1}$ grating and the MIT2 CCD detector were used. The
CCD was of format 4k$\times$2k with 15$\mu$m pixels, which results in a
1$''$ entrance slit projecting onto four pixels on the detector. To avoid 
this oversampling and increase the signal-to-noise per resolution element, 
we binned the CCD on readout to 2k$\times$1k. Two wavelength regions were
observed, one centred on 4335\AA\, and the other on 6560\AA. The slit
width was kept constant at 1'', and the FWHM of the arc lines at the centres
of these settings were measured at 0.11\AA\ and 0.17\AA\ respectively. One 
exposure of 1200s was taken in the red setting, and 4$\times$1200s were
taken in the blue; exposures of comparison arc lamps were taken 
immediately after each science set of exposures. 
The purpose of the red setting was mainly to get 
the H$\alpha$ profile, while the blue covers the higher lines in 
the Balmer series, and all of the He\,{\sc i} and weaker metal lines 
required for atmospheric diagnostics. The 2D frames were bias corrected, 
flat-fielded, and the spectral orders were optimally extracted using standard 
routines within {\sc iraf}. 
After wavelength calibration the four blue 
frames were cross-correlated (taking the best S/N spectrum as a reference)
to check for velocity shifts. None were found at a significance greater than
the estimated error of $\sim$6\,\kms, so the spectra were co-added. 
The final blue spectra had a dispersion of 0.06\AA\,pix$^{-1}$ in the 
central order, which considerably over-samples the metal line profile 
widths. The mean FWHM of several of the stronger metal lines (in the region
4550-4620\AA) was measured at 1.5\AA\ (98 \kms), hence the spectra were
rebinned to a dispersion of 0.3\AA\,pix$^{-1}$. At the furthest useful 
blue wavelength (3915\AA) this corresponds to 4 resolution elements
sampling the FWHM which is adequate for line profile fitting and significantly
increased the S/N in the continuum. The MIT2 CCD is too large for the 
window of the UCLES camera, hence the ends of each \'{e}chelle order are 
severely vignetted. The ends of each order were thus trimmed, and the 
final useful spectrum has almost complete coverage of 3915$-$5280\AA\ across
15 orders, with gaps of 5-15\AA\ appearing only between the last 5. 
The signal-noise-ratio ranges between 50 at the edge orders to 70 in
the central few orders.

\begin{table*}
\caption[]{Equivalent widths for metal lines of Sher\,25 and comparison stars given in m\AA\ units.}
\label{equiv_widths}
\vspace{0.1cm}
\begin{tabular}{rrrrr}
\hline\hline
Ion and Line       & HD2905 &  HD13854 & HD14956 & Sher\,25\\
\hline
C {\sc ii}  4267.02 & 111    &  155     & 128     & 125 \\
{\tiny \&}  4267.27 & & & \\
C {\sc ii}  6578.10 & ...    &  135     & 105     & 110 \\
 & & & \\

N {\sc ii}  3995.00 & 101    &  308     & 443     & 484 \\
N {\sc ii}  4035.08 & 25     &  89      & 101     & 95  \\
N {\sc ii}  4236.86 & 28     &  78      & 63      & 75  \\
{\tiny \&}  4236.98 & & & \\
N {\sc ii}  4241.18 & 53     &  119     & 88      & 165 \\
{\tiny \&}  4241.78 & & & \\
N {\sc ii}  4447.03 & 70     &  106     & 172     & 134 \\
N {\sc ii}  4607.16 & 21     &  134     & 247     & 265 \\
N {\sc ii}  4613.87 & 16     &  91      & 188     & 160 \\
N {\sc ii}  4630.54 & 79     &  301     & 456     & 398 \\
N {\sc ii}  4803.29 & ...    &  122     & 187     & 107 \\
 & & & \\

O {\sc ii}  4069.62 & 249    &  235     & 171     & 167 \\
 {\tiny \&} 4069.89 & & & \\
O {\sc ii}  4072.16 & 213    &  220     & 190     & 122 \\
O {\sc ii}  4075.86 & 235    &  241     & 204     & 181 \\
O {\sc ii}  4132.80 & 85     &  81      & 63      & 65  \\
O {\sc ii}  4317.14 & 204    &  192     & 135     & 177 \\
O {\sc ii}  4319.63 & 205    &  210     & 155     & 210 \\
{\tiny \&}  4319.93 & & & \\
O {\sc ii}  4345.56 & 264    &  202     & 147     & 184 \\
O {\sc ii}  4347.42 & ...    &  112     &  77     &  91 \\
O {\sc ii}  4349.43 & 397    &  347     & 267     & 297 \\
O {\sc ii}  4351.26 & 183    &  127     & 85      & 103 \\
{\tiny \&}  4351.50 & & & \\
O {\sc ii}  4366.89 & 236    &  217     & 170     & 235 \\
O {\sc ii}  4414.90 & 303    & 338      & 257     & 291 \\
O {\sc ii}  4416.97 & 240    & 270      & 211     & 185 \\
O {\sc ii}  4590.97 & 207    & 173      & 147     & 170 \\
O {\sc ii}  4595.96 & 177    & 137      & 159     & 134 \\
{\tiny \&}  4596.18 & & & \\
O {\sc ii}  4638.86 & 242    & 212      & 126     & 159 \\
O {\sc ii}  4661.63 & 259    & 233      & 202     & 194 \\
O {\sc ii}  4673.74 & 88     & 57       & 53      & 60  \\
O {\sc ii}  4676.24 & 234    & 222      & 150     & 253 \\
 & & & \\
Mg {\sc ii} 4481.13 & 170    & 244      & 233     & 205 \\
{\tiny \&}  4481.33 & & & \\
 & & & \\
Si {\sc iv}  4116.10 & 202   & 141      & 111     & 218 \\
Si {\sc iii} 4552.62 & 397   & 455      & 485     & 555 \\
Si {\sc iii} 4568.28 & 348   & 390      & 438     & 437 \\
Si {\sc iii} 4575.22 & 219   & 241      & 293     & 266 \\
Si {\sc iii} 4813.30 & 50    & 65       & 55      & 49  \\
Si {\sc iii} 4819.72 & 80    & 90       & 100     & 96  \\
Si {\sc iii} 4829.96 & ...   & 90       & 85      & 91  \\
 & & & \\
S  {\sc iii} 4361.53 & 206   & 50       & 60      & 119 \\
\\
\hline
\end{tabular}
\end{table*}

The spectra were normalised by fitting cubic splines to continuum regions
free from absorption lines in {\sc iraf} and afterwards  
were transferred to {\sc starlink} spectral analysis 
program {\sc dipso} (Howarth et al. \cite{How98}). The equivalent widths
of the metal lines and non-diffuse lines of neutral helium were measured by
the non-linear least squares fitting of single or multiple Gaussian
profiles to the normalised spectra. The hydrogen and
diffuse helium lines were not measured in this manner, but the
normalised profiles were extracted directly for comparison with
Galactic standards and with theoretical profiles. 

In order to compare the atmospheric parameters, photospheric abundances
and wind parameters with other typical B-type supergiants, three bright 
stars of similar spectral type (lying within the solar neighbourhood)
were chosen as comparison standards. These were HD2905, HD\,14956 and 
HD\,13854, which 
have very similar H$\epsilon$,$\delta$,$\gamma$ lines and 
Si\,{\sc iv}/Si\,{\sc iii} line strength ratios as Sher\,25. These stars
have been observed with the single-order spectrograph ISIS 
on the WHT, and the spectra have a coverage of 3900-4735\AA\, covering 
the main diagnostic lines of interest (see Smartt et al. 2001 where the 
details of the instrument setup is described). The S/N of these spectra are
very high (in excess of 200) and the line profiles are well resolved,  
hence they make excellent comparison templates. 
Equivalent widths for these stars
were also measured in a similar method described above. These three stars were
also part of the Lennon et al. (\cite{ldf93}) atlas, which 
covers the region 3950-5000\AA, and the H$\alpha$ region
6260-6870\AA. Line strength measurements in wavelength regions not 
covered by the WHT ISIS spectra (mainly 4735-5000\AA) 
were taken from this source. The final values of equivalent widths
for each star can be found in Table\,2.

\begin{table}[h]
\caption{The atmospheric parameters derived for Sher\,25 and the three
comparison supergiants, with the homogeneous method described in 
Sect.\ref{atmosphere}.}
\label{atmos_params}
\vspace{ 0.1 cm}
\begin{tabular}{llllll}\hline
\vspace{ 0.1 cm}
Star    &   & $T_{\rm eff}$ & $\log g$ & $v^{\rm Si}_{\rm turb}$ & $v\sin i$ \\
        &   & (K)   & (cgs)    & (kms$^{-1}$)& (kms$^{-1}$) \\\hline
Sher 25  &  & 22\,300 $\pm$1000 & 2.60 $\pm$0.1 & 15 $\pm5$ & 65  \\
HD2905   &  & 23\,500 $\pm$1000 & 2.70 $\pm$0.1 & 11 $\pm5$ & 80 \\
HD13854  &  & 22\,000 $\pm$1000 & 2.65 $\pm$0.1 & 13 $\pm5$ & 80 \\
HD14956  &  & 21\,250 $\pm$1000 & 2.55 $\pm$0.1 & 12 $\pm5$ & 75 \\
\hline
\end{tabular}
\end{table}

\begin{table*}
\caption[]{Chemical abundances calculated in the model atmosphere analysis for Sher\,25 and
the comparison stars. The absolute abundance derived for each line in each star is given, and 
also the differential abundance of Sher\,25 minus each comparison star.}
\label{chem_abun}
\vspace{ 0.1 cm}
\begin{tabular}{cclllllll}\hline\hline
& & & & & & \multicolumn{3}{c}{\scriptsize Differential Abundance of}\\
& & & & & & \multicolumn{3}{c}{\scriptsize Sher 25 with respect to}\\\cline{7-9}
\small
Element/Ion & Wavelength \AA  & HD2905  & HD 13854 & HD 14956 & Sher 25 & HD2905 & HD 13854 & HD 14956\\\hline
\scriptsize
\\
C \,{\sc ii} & 4267.02  & 7.05 & 7.27 & 7.11 & 7.07  & $+$0.02 & $-$0.20 & $-$0.04\\
& {\tiny \&}  4267.27\,\,\,\,\,\\
C \,{\sc ii} & 6578.10  & ...  & 7.12 & 6.66 & 6.95  & ...     & $-$0.17 & +0.29\\
\\
N \,{\sc ii} & 3995.00  & 7.00 & 7.93 & 8.61 & 8.19  & $+$1.19 & +0.26   & $-$0.42\\
N \,{\sc ii} & 4035.08  & 7.71 & 8.32 & 8.51 & 8.40  & $+$0.69 & +0.08   & $-$0.11\\
N \,{\sc ii} & 4236.98  & 7.31 & 7.86 & 7.78 & 7.84  & $+$0.53 & $-$0.02 & +0.06  \\
N \,{\sc ii} & 4241.18  & 7.64 & 8.07 & 7.93 & 8.28  & $+$0.64 & +0.21 & +0.35\\
& {\tiny \&} 4241.78\,\,\,\,\,\\
N \,{\sc ii} & 4447.03  & 7.20 & 7.53 & 7.98 & 7.45  & $+$0.25 & $-$0.08 & $-$0.53\\
N \,{\sc ii} & 4607.16  & 7.04 & 8.19 & 9.15 & 8.91  & $+$1.87 & +0.72 & $-$0.24\\
N \,{\sc ii} & 4613.87  & 7.04 & 8.02 & 8.90 & 8.45  & $+$1.41 & +0.43 &  $-$0.45\\
N \,{\sc ii} & 4630.54  & 7.10 & 8.31 & 9.42 & 8.53  & $+$1.43 & +0.22 & $-$0.89\\
\\
O \,{\sc ii} & 4069.62  & 8.95 & 8.91 & 8.65 & 8.39  & $-$0.56 & $-$0.52 & $-$0.26\\
& {\tiny \&}   4069.89\,\,\,\,\,\\
O \,{\sc ii} & 4072.16  & 9.03 & 9.10 & 9.03 & 8.20  & $-$0.83 & $-$0.90 & $-$0.83\\
O \,{\sc ii} & 4075.86  & 8.99 & 8.98 & 8.88 & 8.37  & $-$0.62 & $-$0.61 & $-$0.51 \\
O \,{\sc ii} & 4132.80  & 9.04 & 9.03 & 8.93 & 8.78  & $-$0.24 & $-$0.25 & $-$0.15\\
O \,{\sc ii} & 4153.30  & ...  & 9.29 & 8.95 & 8.95  & ...     & $-$0.33 & +0.00\\
O \,{\sc ii} & 4317.14  & 9.13 & 9.06 & 8.76 & 8.87  & $-$0.26 & $-$0.19 & +0.00\\
O \,{\sc ii} & 4319.63  & 9.06 & 9.12 & 8.87 & 8.91  & $-$0.15 & $-$0.21 & +0.04\\
& {\tiny \&}  4319.93\,\,\,\,\,\\
O \,{\sc ii} & 4366.89  & 9.22 & 9.08 & 8.90 & 8.97  & $-$0.25 & $-$0.11 & $+$0.07\\
O \,{\sc ii} & 4590.97  & 9.28 & 9.04 & 9.03 & 8.88  & $-$0.40 & $-$0.16 & $-$0.15\\
O \,{\sc ii} & 4595.96  & 9.19 & 8.91 & 9.24 & 8.80  & $-$0.39 & $-$0.11 & $-$0.44\\
& {\tiny \&}  4596.18\,\,\,\,\,\\
O \,{\sc ii} & 4638.86  & 9.44 & 9.45 & 9.19 & 9.37  & $-$0.07 & $-$0.08 & +0.18\\
O \,{\sc ii} & 4661.63  & 9.35 & 9.21 & 9.17 & 8.75  & $-$0.60 & $-$0.46 & $-$0.42\\
O \,{\sc ii} & 4673.74  & 9.13 & 8.83 & 8.86 & 8.80  & $-$0.33 & $-$0.03 & $-$0.06\\
O \,{\sc ii} & 4676.24  & 9.28 & 9.23 & 8.85 & 8.95  & $-$0.33 & $-$0.28 & $+$0.10\\
\\
Mg \,{\sc ii} & 4481.13 & 7.44 & 7.58 & 7.54 & 7.46  & $-$0.02 & $-$0.12 & $-$0.08\\
& {\tiny \&} 4481.33\,\,\,\,\,\\
\\
Si \,{\sc iii} & 4552.62  & 7.90 & 8.00 & 8.43 & 8.17  & $+$0.27 & +0.17 & $-$0.26\\
Si \,{\sc iii} & 4568.28  & 7.93 & 7.97 & 8.50 & 8.00  & $+$0.07 & +0.03 & $-$0.50\\
Si \,{\sc iii} & 4575.22  & 7.88 & 7.87 & 8.40 & 7.92  & $+$0.04 & +0.05 & $-$0.48\\
\\
Si \,{\sc iii} & 4813.30  & 7.35 & 7.45 & 7.49 & 7.28  & $-$0.07 & $-$0.17 & $-$0.21\\
Si \,{\sc iii} & 4819.72  & 7.46 & 7.54 & 7.81 & 7.54  & $+$0.08 & +0.00 & $-$0.27\\
Si \,{\sc iii} & 4829.96  & ...  & 7.40 & 7.53 & 7.37  & ...     & $-$0.03 & $-$0.16\\
\\
Si \,{\sc iv} & 4116.10  & 7.22 & 7.41 & 7.56 & 7.43   & $+$0.21 & +0.02 & $-$0.13\\
\\
\hline\hline
\end{tabular}
\end{table*}

\begin{table*}
\caption[]{
The mean of the individual line abundances from Table\,\ref{chem_abun}
for each element in each star. Also listed is the mean of the
differential results. The absolute values are quoted on the
logarithmic scale $12 + log [N_x]/[N_{\rm H}]$ (see
Sect.\,\ref{photosphere}). The mean abundances were determined by
first determining the fractional abundance $[N_x]/[N_{\rm H}]$ for each
line, calculating the linear mean and quoting this on the original
logarithmic scale. The errors quoted are the standard errors in the
mean ($\sigma/\sqrt{n}$); where $n$ is the number of features used,
and $\sigma$ is the standard deviation of the sample again calculated
on a linear scale. These represent the statistical scatter of the
measurements, however a further systematic error will be present in
each case.}
\label{mean_abun}
\vspace{ 0.1 cm}
\begin{tabular}{llllllll}\hline
 & & & & & \multicolumn{3}{c}{\scriptsize Differential Abundance of}\\
 & & & & & \multicolumn{3}{c}{\scriptsize Sher 25 with respect to}\\\cline{6-8}
\small
 Ion  & HD2905 &  HD13854 & HD 14956 & Sher 25 & HD2905 & HD 13854 & HD 14956\\\hline
\\
C\,{\sc ii} &   7.05          & 7.20$\pm$0.07 & 6.94$\pm$0.17 &  7.01$\pm$0.06 & $+$0.02            & $-$0.18$\pm$0.01 &   +0.16 $\pm$0.13 \\
N\,{\sc ii} &   7.34$\pm$0.10 & 8.09$\pm$0.07 & 8.70$\pm$0.13 &  8.42$\pm$0.12 & $+$1.29$\pm0.16$   &   +0.30$\pm$0.10 & $-$0.13 $\pm$0.12 \\
O\,{\sc ii} &   9.18$\pm$0.14 & 9.12$\pm$0.05 & 8.98$\pm$0.04 &  8.87$\pm$0.07 & $-$0.34$\pm0.16$   & $-$0.25$\pm$0.05 & $-$0.10 $\pm$0.06 \\
Mg \,{\sc ii} & 7.44          & 7.58          & 7.54          &  7.46          & $-$0.02            & $-$0.12          & $-$0.08           \\
Si\,{\sc iii} & 7.41$\pm$0.05 & 7.47$\pm$0.04 & 7.63$\pm$0.10 &  7.41$\pm$0.07 & $-$0.00$\pm0.05$   & $-$0.06$\pm$0.05 & $-$0.21 $\pm$0.03 \\
Si \,{\sc iv} & 7.22          & 7.41          & 7.56          &  7.43          & $+$0.21            &   +0.02          & $-$0.13           \\
\\
\hline
\end{tabular}
\end{table*}

\section{Model atmosphere and wind analysis}
\label{atmosphere}

\subsection{Photospheric analysis and abundances}
The photosphere of Sher\,25, and subsequently the three comparison stars,
were modeled using the techniques described in detail in Paper\,I
and briefly summarized in Smartt et al.
(\cite{sma2001a}). A grid of non-LTE model atmospheres has previously
been generated using the code {\sc tlusty} (Hubeny \cite{hub88}) 
covering an effective temperature range $10\,000 \leq T_{\rm eff} \leq 
35\,000$ and gravities from $\log g = 4.5$ to the Eddington stability 
limit at each $T_{\rm eff}$. Increments were 2500\,K and 0.25\,dex
in $T_{\rm eff}$ and $\log g$ respectively. Models were 
calculated for two helium fractions, i.e. $y=0.09$, and 
$y=0.20$, where $y=N[He]/N[H+He]$. 
The models contain only H and He and assume a plane-parallel atmosphere
which is in hydrostatic equilibrium. The consequences of these 
assumptions have been discussed in Paper\,I. 
Line formation calculations for all the ionic species with 
absorption lines observed in the spectra of Sher\,25 were 
calculated using the codes {\sc detail} (Giddings \cite{gid81}), 
and {\sc surface} (Butler \cite{but84}). Microturbulence 
was included in the calculations as an extra free parameter, and 
the statistical equilibrium calculations were performed for the
species H\,{\sc i}, He\,{\sc i}, C\,{\sc ii}, N\,{\sc ii},
O\,{\sc ii}, Si\,{\sc iii}, Si\,{\sc iv} and Mg\,{\sc ii}. The
atomic data employed were identical to those used in 
Paper\,I.

Although the photospheric analysis is based on NLTE atmospheres, they
are plane-parallel and do not account for metal line blanketing or 
wind contamination. In this paper we concentrate on differential 
abundance determinations and comparing stars of similar atmospheric
parameters, and acknowledge that the absence of these physical 
effects in our models may compromise the final absolute abundances. 
In the future it is our desire to pursue a fully unified model atmosphere
and wind analysis of sher25 with even higher quality spectral data
from, for example, the UVES spectrometer on the VLT. This would allow an
excellent comparison between the very weak metal lines formed deep in the
photosphere, and the stronger lines we see in the current ucles spectra.

\subsubsection{Determination of atmospheric parameters}
\label{params_method}
The relative strengths of two ionization stages of silicon can be used to
determine the effective temperature for stars of early-B spectral types. 
Absorption lines of Si\,{\sc iii} and Si\,{\sc iv} are well observed and 
reliably measured in the UCLES spectra of Sher\,25. The line strengths
are listed in Table\,\ref{equiv_widths}, with six lines of 
Si\,{\sc iii} and one of Si\,{\sc iv} initially used to determine 
$T_{\rm eff}$. There is an inherent problem in
{\sc detail} when calculating the Si\,{\sc iii} 
line strengths of the 4552\AA\ triplet, which has been discussed in 
Paper\,I. In the lowest gravity models 
having effective temperatures of 22\,500K or higher, {\sc detail}
calculations showed an overpopulation of the upper levels 
of Si\,{\sc iii} in atmospheric regions where the lines are 
formed. This is particularly problematic for the 4552\AA\ triplet
but is less pronounced for the 4813\AA\ triplet which is formed
deeper in the atmosphere. The effect leads to a filling-in of 
the 4552\AA\ multiplet line profiles, and hence one needs to increase
the Si abundance substantially to offset this effect. One can 
readily see this in Table\,\ref{chem_abun}, were the mean 
abundances derived from the 4552\AA\ triplet are between 0.5$-$0.8\,dex
higher than those derived from 4813\AA. We follow the 
conclusion in Paper\,I that the 4552\AA\ lines are not
reliably modeled with our methods and use only the 4813\AA\
triplet in determining the effective temperature of all four 
stars.

The shape of the Balmer line profiles (H$\epsilon$, H$\delta$ and H$\gamma$, 
and H$\beta$) are useful indicators of the stellar surface gravity. 
In particular the width of the line wings are excellent indicators of 
$\log g$. Assuming an initial \teff, the surface gravity is calculated by
fitting the observed profiles with theoretical lines calculated at
different  values of $\log g$, paying particular attention to the line 
wings. The mean of the results from H$\delta$ and H$\gamma$ was used. 
H$\epsilon$ suffers from interstellar contamination (Ca\,{\sc ii})
and a stellar blend with He\,{\sc i}. Both H$\beta$ and H$\alpha$ can show
significant emission contributions from the stellar wind and 
the latter is discussed in Sect.\ref{winds}.
Some iteration in the simultaneous determination of \teff\ and $\log g$
was necessary, given the interdependence of each parameter but 
convergence was normally achieved within 3-4 cycles. 

The microturbulent velocity was estimated by requiring that each line
of the Si\,{\sc iii} triplet at 4552\AA\ gave the same abundance. 
The determination of \vt  ~in this way has been discussed in 
Paper\,I and it has the advantage that three lines from the same multiplet
with quite different equivalent widths are used. The lines of 
O\,{\sc ii} can also be used and these often give higher values 
compared to those derived when using Si\,{\sc iii}. 
Due to the homogeneity of the atomic data and the large range in 
line strengths covered,  we use the value derived from 
the Si lines, and note that as we 
re-derive all parameters for the standards (and subsequent abundances) 
in a homogeneous way any systematic error in this procedure will be 
mirrored in Sher\,25 and the three comparison stars. 

\begin{figure*}
\label{Hline_spec}

\epsfig{file=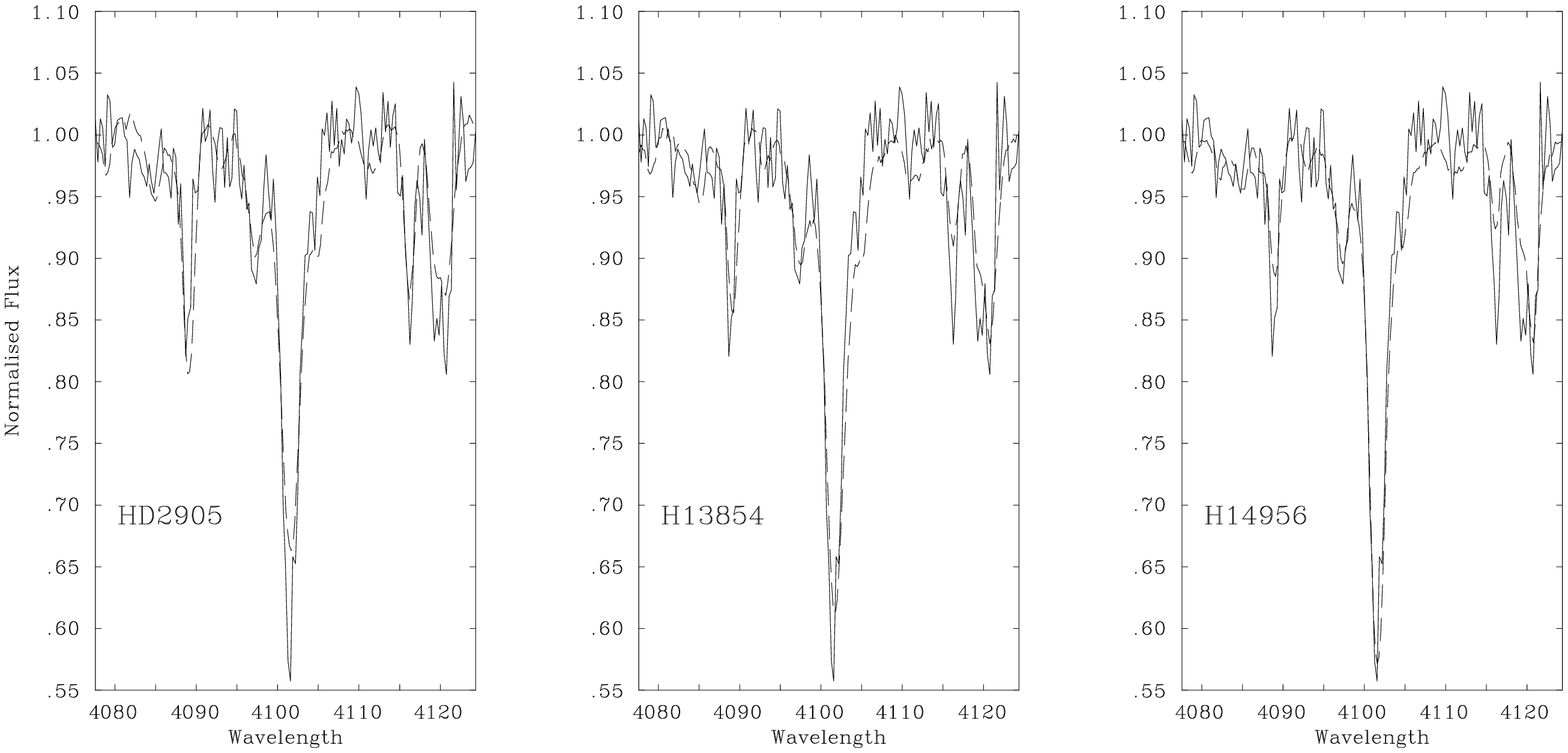, width=10cm,angle=270}

\epsfig{file=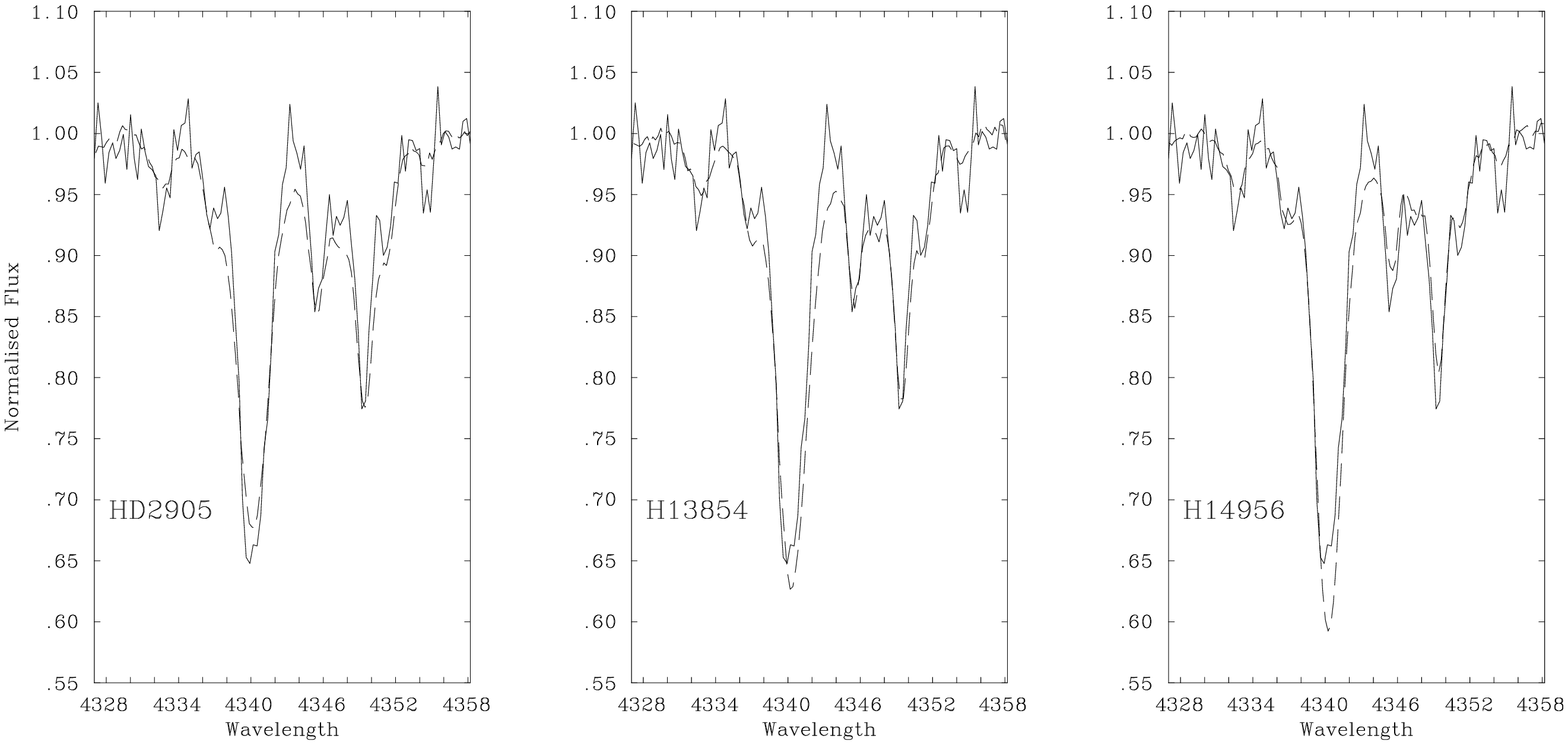, width=10cm,angle=270}

\caption[]{The Balmer line profiles of H$\delta$ and H$\gamma$
for Sher\,25 and the three comparison stars. In each case the Sher\,25
spectrum is the solid line, and the dashed profile is that from the 
three stars as labeled in each panel.}
\end{figure*}

\begin{figure*}
\label{Heline_spec}

\epsfig{file=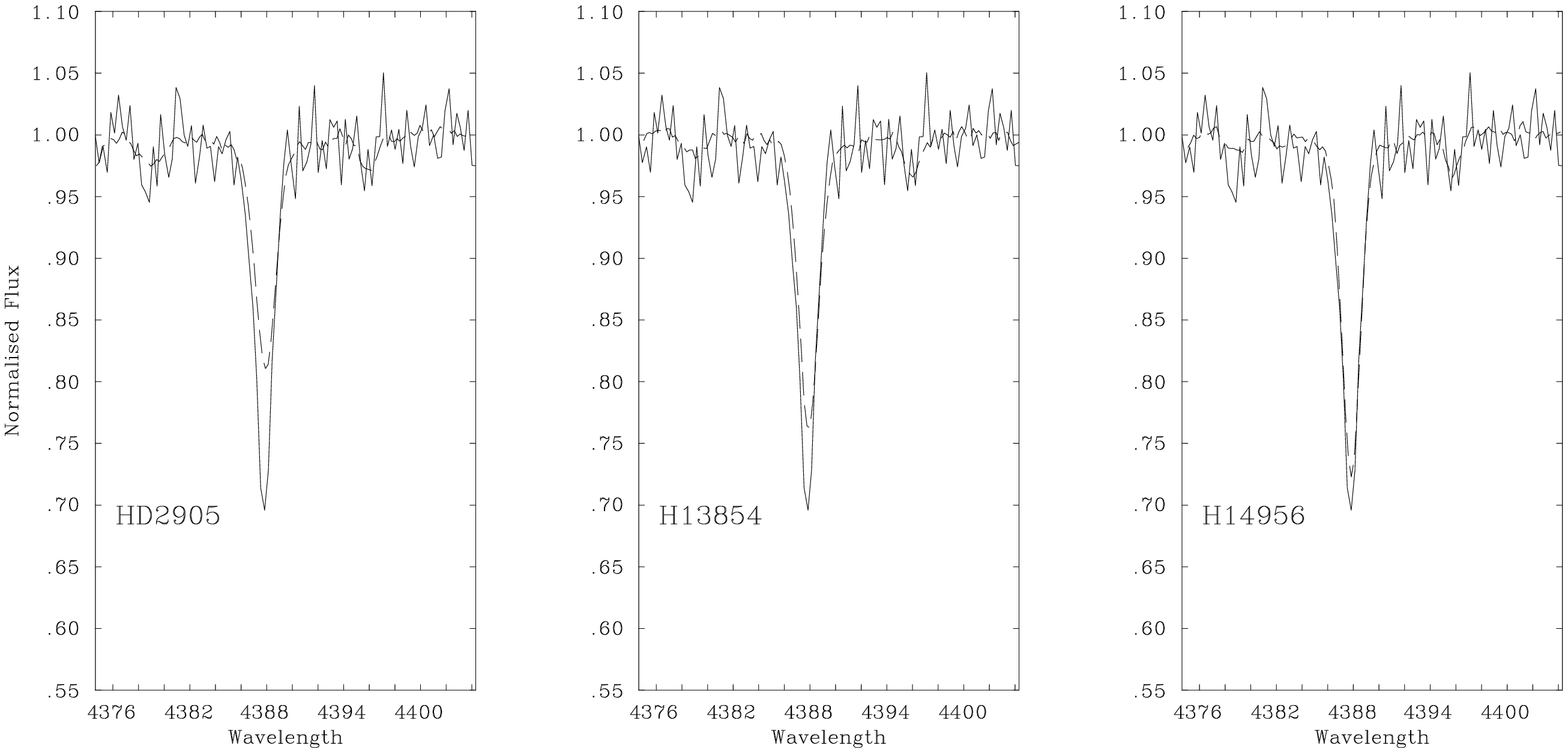,width=10cm,angle=270}

\epsfig{file=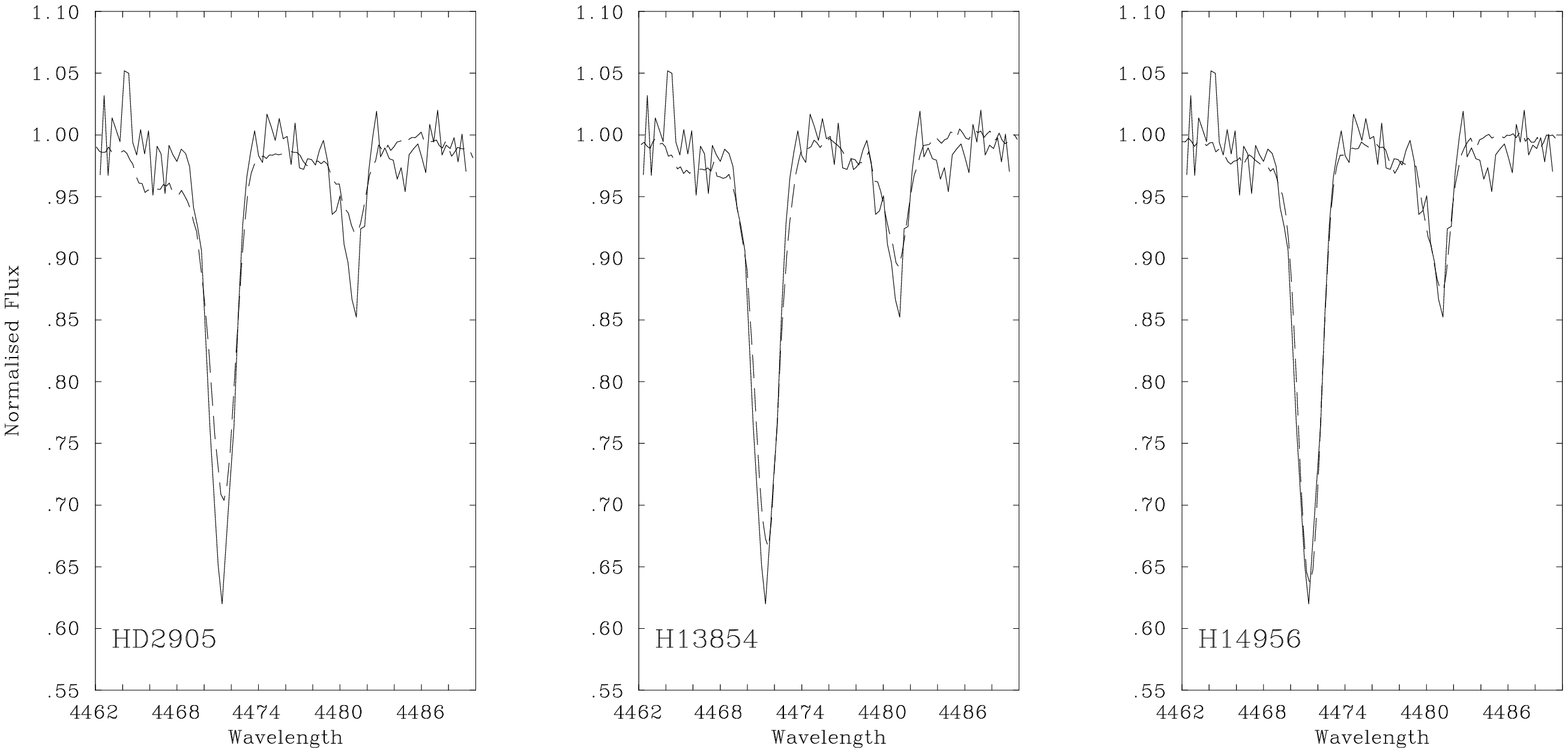,width=10cm,angle=270}

\caption[]{The line profiles of He\,{\sc i} 4471\AA\ and 
He\,{\sc i} 4387\AA\ 
for Sher\,25 and the three comparison stars. In each case the Sher\,25
spectrum is the solid line, and the dashed profile is that from the 
three stars as labeled in each panel.}
\end{figure*}

The three parameters of \teff, $\log g$ and \vt\ were varied so that a
simultaneous fit to all three was achieved for Sher\,25, and the final
atmospheric parameters are listed in Table\,\ref{atmos_params}. 
The $v\sin i$ of the star was determined by convolving the 
theoretical line profiles of prominent metal lines with a rotational
broadening function until good agreement with the observed profile
was achieved (as discussed in Lennon et al. \cite{len91}). 
We then chose three stars which had very similar atmospheric 
parameters from the list of bright nearby B-type supergiants
analysed in Paper\,I. Two stars of almost identical
\teff\ are HD14956 and HD13854, and the observed
profiles of the hydrogen and helium lines were compared directly
to make sure the luminosity class and the helium line strengths
(and hence the atmospheric helium composition) were indeed very
similar. This can be seen in Figs. \ref{Hline_spec} and 
\ref{Heline_spec}. A third star HD2905 was also chosen as an 
extra comparison due to its relatively weak nitrogen line spectra,
hence giving us comparison stars which are both qualitatively 
nitrogen rich and poor. 

Using the higher quality spectra now available to us (as described in
Sect.\,\ref{obsdata}) we have re-derived the atmospheric parameters of
the comparison stars using identical methods to that used for
Sher\,25.  Our results for HD\,2905 and HD\,14956 are identical to those in
Paper\,I, however we derive a \teff\ which is 1500\,K
cooler than the previous analysis for HD\,13854. 
The reason for this is likely to be
the slightly different equivalent widths of the Si\,{\sc iv} 4116\AA\ line
measured in each study. 
We measure this line as having a strength of 111m\AA\, and it 
is quoted as 160m\AA\ in the spectra 
of Lennon et al. (1993). In
order to keep this analysis as consistent as possible we use our
refined temperature, which is
similar to the previous result in Paper\,I 
within the quoted errors.
The line profiles and strengths of the He\,{\sc i} lines were all
similar in Sher\,25 and the 3 comparisons (as seen in 
Fig.\,\ref{Heline_spec}). This suggests a very similar He abundance
across the four stars within the expected uncertainties 
however there is difficulty in modeling all the singlet and triplet
lines simultaneously and satisfactorily (see Paper\,I). 
The lines which appear to be modeled satisfactorily
are the singlets He\,{\sc i}  4437\AA\ and 4387\AA\, which 
form deepest in the atmosphere. However the  4437\AA\ line 
falls on top of the diffuse interstellar band at 4430\AA\, which is 
particularly strong in the Sher\,25 sight line, and does not allow 
accurate extraction of the stellar line. 
The 4387\AA\ line in Sher\,25 gives a helium  
fraction $y=0.1\pm0.02$, hence there is no evidence of 
a He overabundance. The 4471\AA\ line, being a triplet, is difficult to 
model in quantitative terms 
but the similarity in the empirical spectra of all four supergiants, 
and a measured equivalent width of 820\AA\ (compare with Fig.\,3 in
Paper\,I) suggests that there is no evidence for using a non-solar
He value in the Sher\,25 model atmospheres and that all four stars
have comparable He abundances. 

\begin{figure}[h]
\epsfig{file=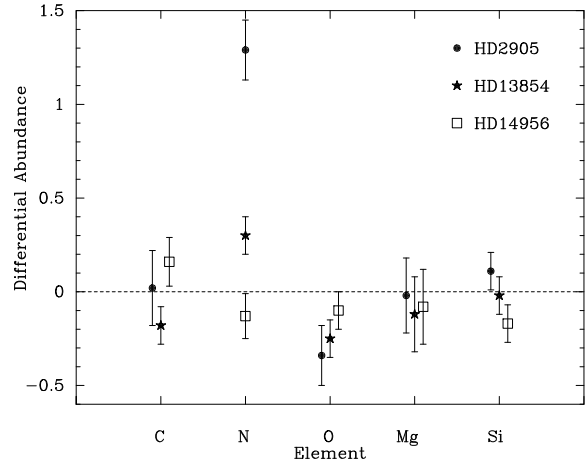,width=9cm}
\caption[]{Differential abundance of Sher\,25 with respect to HD2905,
HD14956 and HD13854. The values and errors are taken from 
Table\,\ref{mean_abun}, with the minimum error assumed to be 
0.1\,dex.}
\label{diff_abun}
\end{figure}

\subsubsection{Photospheric abundances}
\label{photosphere}

The atmospheric parameters were used to define final model
photospheres for all four stars. All unblended metal-line absorption
features identified in Sher\,25 (and the same features in the other
stars) were modeled using the line formation codes {\sc detail} and
{\sc surface} as described in Sect.\,\ref{params_method}.  A curve of
growth method was used to determine an ionic abundance for each line,
and the results are quoted on the scale $12 + log [N_x]/[N_{\rm H}]$
where $N_x$ and $N_{\rm H}$ are the abundance of ion $x$ and hydrogen
by number (see Table\,\ref{chem_abun}).  The mean of all the
individual lines were calculated, by first determining $N_x/N_{\rm H}$,
calculating the mean and then converting back to the logarithmic
scale. Standard deviations ($\sigma$) and standard errors in the mean
($\sigma/\sqrt{n}$; where $n$ is the number of features used) were
calculated and are quoted in Table\,\ref{mean_abun}.  The differential
abundance of Sher\,25 with respect to each comparison star was done on
a line-by-line basis and the mean values of these results were
calculated in a similar way. These results should be more reliable
than the absolute abundances as the differential approach mitigates
the uncertainties in the atomic data and physical methods. We discuss
the results for each element briefly below, and the implications of
these will be discussed in Sect.\,\ref{discussion}.

{\em Carbon:} The carbon abundance is based on two lines only, being the
only reliably measured features in the Sher\,25 spectrum. There is a 
C\,{\sc ii} doublet at 6578.10\AA\ and 6582.85\AA, however the latter 
is contaminated by the nebular [N\,{\sc ii}] 6583.6\AA\ line. There
is quite strong nebulosity around Sher\,25 which was clearly visible
on the raw 2D CCD images, and perfect subtraction of the 
features was not possible during data reduction. Hence the 6582.85\AA\
line of C\,{\sc ii} was not usable, although it is reliably measured
in the bright comparison stars. The 
4267\AA\ doublet is well known to give unreliable absolute
abundances with the particular line formation codes we have used 
(e.g. Vrancken et al. \cite{vra2000}, Lennon \cite{len83}, Paper\,I), 
hence the values from this feature should be treated with 
considerable caution. Even in B-type dwarfs this line has historically
proved difficult to model reliably (in either LTE or NLTE) and 
Sigut (\cite{sigut96}) suggests that the model atom of
Eber \& Butler (\cite{ebut88}) which we have employed needs extension
to improve results from this line. 
The differential abundances should be more reliable, and indicate that
there appears little difference, within the errors, in the carbon
abundance in the four stars.  The absolute abundance derived in all
four stars is very low compared to values derived in solar neighbourhood
B-type stars e.g. 8.2 from Kilian (\cite{kil92}) and Gies \& Lambert 
(\cite{gies92}) which suggests that our absolute values cannot be 
relied upon to be used as a quantitative test of evolutionary 
models (see Sect.\,\ref{discussion}). The only other line
in common with all stars (6578\AA) gives similar  results to 4267\AA, 
but Paper\,I found that this was only satisfactorily modeled
at temperatures below 20000K, and that the hotter models 
have this line in emission. Hence with our current codes, 
we do not have a good measure of the absolute carbon abundance in
these stars, although we can say that the abundances appear
fairly similar in each atmosphere. 

{\em Nitrogen:} Eight lines of N\,{\sc ii} were identified and measured
in Sher\,25 and each were found in the comparison stars, hence 
giving a good measurement of the nitrogen abundance in statistical terms. 
As expected, the nitrogen abundance in HD2905 is much lower than in 
either of the other two comparison stars and in Sher\,25. This 
star has very weak N\,{\sc ii} absorption features and has
been classified as a BC0.7Ia (Lennon et al. 1992). This star was
chosen as a comparison to show the difference between the nearby 
B-type supergiants that is observed. The nitrogen abundance in 
Sher\,25 is much greater than in HD2905, and is intermediate
between the other two stars (illustrated in Fig\,3). 
Paper\,I indicates that at the 
$T_{\rm eff}$ of our stars, the N\,{\sc ii} line strengths in the 
bulk of the stars are reasonably well modeled with a ``normal'' 
nitrogen abundance of 7.7\,dex. Problems with overpopulation
of upper atomic levels and emission features appearing in 
line profiles were not in evidence for N\,{\sc ii} in this 
temperature regime, unlike C\,{\sc ii} and Si\,{\sc iii}. 
Hence we have more confidence in using the absolute values of 
N\,{\sc ii} as being useful for quantitative comparisons with 
evolutionary calculations discussed in Sect.\,\ref{discussion}.

{\em Oxygen, magnesium and silicon:} With fourteen lines measured in Sher\,25 
oxygen is the best sampled element, as is normal in early B-type stars. 
Paper\,I found that the O\,{\sc ii} absolute 
abundances in this temperature range did not suffer any obvious problems
with high photoionization rates. Given the good statistical sample
of O\,{\sc ii} features, and the encouraging analysis in Paper\,I we
assume that the oxygen abundance derived in our stars is reliable enough 
to use in comparison with evolutionary calculations. 
There is some evidence that the oxygen abundance in 
Sher\,25 is somewhat lower than in the three standard stars, in particular
in comparison with HD2905 and  is discussed in Section\,\ref{discussion} below
where we consider the implications of CNO processing in these objects.
Both magnesium and silicon in Sher\,25 are very similar to the
abundances derived in the standard stars. As these two elements 
should not be affected by any mixing or dredge-up processes that 
may affect the CNO surface composition, the fact that they are
similar in all three stars points towards them all having fairly 
similar initial chemical compositions. 

\begin{table}
\caption{The errors listed in this table illustrate how the abundances in Sher 25 
would vary within the error bars of the atmospheric parameters. 
The values were calculated around the model atmosphere with
 $\log  g$ \,= 2.6, $T_{eff}$\,=\,22\,300, \,$\xi$\,=\,15\,km\,s$^{-1}$.
The final column lists the total error estimated by adding the 
individual errors in quadrature.}
\label{errors}
\vspace{ 0.1 cm}
\begin{tabular}{lcccc}\hline
\vspace{ 0.1 cm}
Species & {$\log g$ $\pm$0.2} &  {$T_{eff}$ $\pm$1\,000}& {$\xi \pm$5} & $\Delta \log {\rm E/H}$ \\\hline
\vspace{ 0.1 cm}
C \,{\sc ii}   & $\pm$0.03  & $\pm$0.02  & $\pm$0.06  & $\pm$0.07 \\
N \,{\sc ii}   & $\pm$0.18  & $\pm$0.10  & $\pm$0.27  & $\pm$0.34 \\
O  \,{\sc ii}  & $\pm$0.13  & $\pm$0.15  & $\pm$0.18  & $\pm$0.18  \\
Mg  \,{\sc ii} & $\pm$0.27  & $\pm$0.15  & $\pm$0.11  & $\pm$0.33 \\
Si \,{\sc iii} & $\pm$0.18  & $\pm$0.10  & $\pm$0.10  & $\pm$0.23 \\
\hline
\end{tabular}
\end{table}

\subsection{Stellar wind analysis}
\label{winds}

As the origin of the gas surrounding Sher\,25 is a key factor in the
evolutionary history of this object, an investigation of the stellar
wind properties is highly desirable. Further we can compare results
directly with the extensive study of Paper II (Kudritzki et al. 1999) which
investigated the wind properties of 14 Galactic early B-type supergiants.
The method used to determine the mass-loss rate for Sher\,25 is 
described in detail in Paper\,II. Exactly the
same code is used in this case (Santolaya-Rey et al. \cite{sph97}) and here
we briefly outline the procedure used. 

\begin{figure}[h]
\epsfig{file=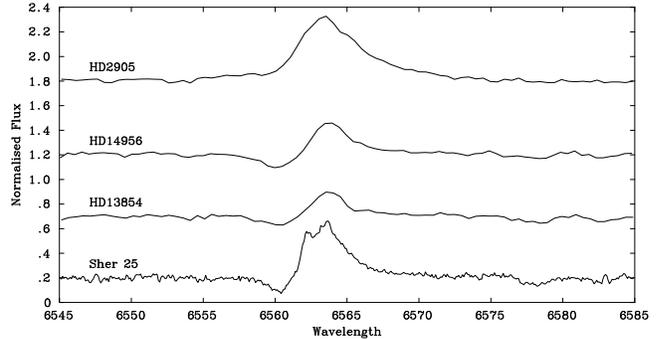,width=9cm}
\caption[]{The H$\alpha$ profiles of Sher\,25 and the three comparison 
stars. The secondary 
emission peak bluewards of the rest wavelength of H$\alpha$ in Sher\,25 is
due to imperfect nebular subtraction in the 2D image. Weak residuals of 
the [N\,{\sc ii}] 6583.6\AA\ \& 6548.1\AA\ nebular lines can also be 
seen. The H$\alpha$ emission of Sher\,25 is similar in strength to the 
comparison stars and the wind analysis implies it has quite normal
mass-loss}
\label{halpha_comp}
\end{figure}

The stellar parameters determined above were used to calculate the 
emergent flux from a non-LTE model atmosphere, and by assuming an 
absolute visual magnitude of the star, a radius can be determined.
However the calculation of absolute V-band magnitude depends on our
knowledge of distance and reddening. The former has been 
calculated by several authors, the most recent of which is
6.3 $\pm$0.6\,kpc (Pandey et al. \cite{pandey2000}). Sher\,25 has 
$B-V=1.4$, which indicates a extinction E(B-V)=1.6 from the 
intrinsic colours of Deutschman et al. (\cite{Deu76}). Pandey et al. also 
determine R (the ratio of total to selective reddening)
towards NGC3603 and suggest a value of 4.3, which is much higher than
the normal Galactic line of sight value R=3.1 (Seaton \cite{sea79}). 
However this may be applicable more toward the central regions
of the cluster, whereas Sher\,25 sits outside the main body of 
young O-stars. We experimented with the value of 4.3 and determined 
what effect it would have on the derived intrinsic luminosity of the
star and its radius and whether this was consistent with the 
observed H$\alpha$ profile. Assuming R=4.3 would imply a luminosity of 
$\log L/L_\odot=6.2$, and a radius of 80R$_\odot$. These values 
suggest the star would be more luminous and have a larger radius
than HD190603 (Paper\,I, Lennon et al. 1992), but the 
surface gravity of Sher\,25 we derive is incompatible with a very
low gravity, extremely high luminosity star. HD190603 is 
probably the most luminous star
in the Paper\,I sample, and shows H$\alpha$ very strongly
in emission. Sher\,25 has weaker H$\alpha$ emission and broader absorption 
wings in the other Balmer lines, which strongly suggests it is a lower
luminosity object. Also a luminosity of 6.2 would suggest that
Sher\,25 is much more luminous, and hence has a much lower
surface gravity than either of HD2905 or HD13854, both of which have
undergone a wind analysis in Paper\,II. However the H$\alpha$ emission
line in Sher\,25 is not significantly stronger than in either of
the latter two, and the Balmer line profiles do not suggest a much
lower surface gravity. Adopting a value of R=4.3 would 
lead to serious inconsistencies between the luminosity of
Sher\,25 and its spectral properties, hence it is very unlikely
that such a high value of R is applicable. There does appear 
to be evidence for a
variation in R toward this sight line, and we adopt a value of R=3.7
as being indicative of the true value with a probable error of $\pm$0.5. 
With this value we obtain a radius of $59 \pm 15 R_\odot$. 
The observed H$\alpha$
profiles of the stars are shown in Fig.\ref{halpha_comp}.
The emission peak on the blue wing of Sher\,25 is indicative of imperfect
nebular subtraction, however the rest of the profile will not be affected
given its width and the narrow nature of the nebular line. 
Crowther \& Dessart (\cite{crow98}) have also derived a distance to 
NGC3603 based on the absolute magnitudes of O-stars. They find a distance
of 10.1\,kpc, using the standard value R=3.1. We find that 
Using these numbers puts 
luminosity and radius of Sher\,25 at very similar values to those
derived using d=6.3\,kpc and R=3.7 i.e. 
$59 \pm 15 R_\odot$ and $\log L/L_\odot=5.9 \pm 0.2$.

\begin{figure}[h]
\epsfig{file=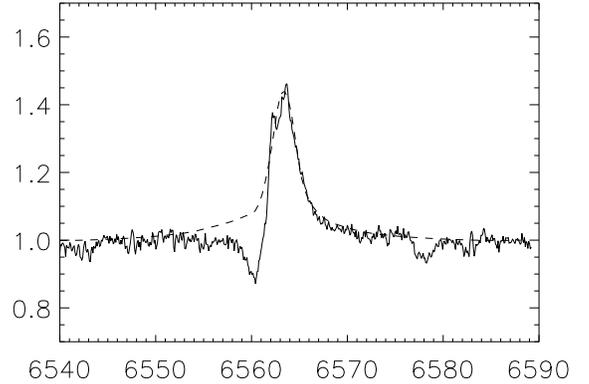,width=9cm}
\caption[]{The H$\alpha$ profile of Sher\,25 and the fit from the 
unified wind models. The peak is 
well reproduced, whereas the bluewards absorption is not seen
in our model spectra due to the reasons described in Sect.\ref{winds}}
\label{halpha_fit}
\end{figure}

No UV spectra exist to determine a terminal velocity for the wind 
of Sher\,25 we adopt a value of 1000\kms. The values of 
$v_{\infty}$ for HD2905 and HD13854 are derived in Paper\,II and values 
of 1100 and 1000\,\kms were found, indicating our adopted value for 
Sher\,25 is reasonable.
The mass-loss rate $\dot{M}$ can be 
obtained from fitting the H$\alpha$ profile, using the non-LTE 
unified model atmosphere technique incorporating the stellar wind 
and spherical extension, as described in Paper\,II.
A grid of models was calculated with different mass-loss rates
and the best fit was found 
for $1.75\times10^{-6}$\msol\,yr$^{-1}$, and $\beta=1.5$
(see Fig.\,\ref{halpha_fit}). The mass-loss rate
is close to (and roughly intermediate between)
 the values  for HD13854 and HD2950 in Paper\,II, and 
appears fairly typical for early-B supergiant stars with solar-type
chemical compositions. The fit to the emission profile is good, but
the absorption dip blueward of the peak is not reproduced. This
dip is also seen in other early B-types of similar atmospheric
parameters and mass-loss rates, and the failure of the wind 
analysis calculations to match the observed shape is discussed
in Paper\,II. It may be due to the neglect of metal-line
blanketing in the models, or wind variations and deviations from 
spherical symmetry. We further derive a luminosity of
$\log L/L_\odot=5.9 \pm 0.2$, and a modified wind momentum of 
$\log (\dot{M}v_{\infty}(R/R_{\odot})^{0.5}) = 28.9 \pm 0.3$, which would
place the star close to the wind momentum-luminosity relationship
for solar metallicity 
early B-type stars (see Paper\,II). Hence the most important 
conclusion from the 
fitting of the H$\alpha$ and derivation of its wind and luminosity
properties is that the wind of Sher\,25 looks fairly typical for an early
B-type supergiant. 

\section{Discussion of the evolutionary scenarios for Sher\,25}
\label{discussion}

\begin{table*}
\caption[]{Comparison of CNO abundances from our analysis of the four
B-type supergiants. For comparison the values for solar neighbourhood
B-type main-sequence stars are taken from
Gies \& Lambert (\cite{gies92}), and the solar values are from Grevesse \& Sauval
\cite{grev98}. The C, N and O element abundances are by number fraction. 
The numbers in italics should be treated with extreme caution, given the
uncertainty in the absolute C abundance as discussed in Sect.\ref{photosphere}.}
\label{CNO_comparison}
\vspace{ 0.1 cm}
\begin{tabular}{lllllll}\hline
\small
 Element & Solar &  B-stars & HD2905 & HD13854 & HD14956 & Sher\,25 \\\hline
 & \multicolumn{6}{c}{Element abundance by number ($\times10^{-3}$)}\\
C &  0.33 &  0.16  & {\em 0.01} & {\em 0.02} & {\em 0.01} & {\em 0.01} \\
N &  0.08 &  0.07  &      0.02  &      0.12  &      0.50  &      0.26 \\
O &  0.68 &  0.48  &      1.51  &      1.32  &      0.95  &      0.74 \\
$\frac{\rm C+N+O}{\rm H+He}$ & 0.99 &  0.65 & 1.41 & 1.33 & 1.33 & 0.92 \\
\\
 & \multicolumn{6}{c}{Elemental fractions}\\
N/C & 0.25  & 0.41  & {\em 2.0}   & {\em 7.7}   & {\em 55.7} & {\em 26.3} \\
N/O & 0.12  & 0.14  & 0.02 & 0.09 & 0.53 & 0.36 \\
\hline
\end{tabular}
\end{table*}

\subsection{A post red-supergiant star in a ``blue-loop'' }

The existence of the ring shaped nebula and bi-polar outflow around Sher\,25
has been interpreted by Brandner et al. (\cite{bran97a}, \cite{bran97b})  
as arising during 
a previous red supergiant evolutionary phase. In this scenario the star
began its life as an O-type main-sequence object, burning H to He 
through the CN and the ON-cycles. After core
hydrogen exhaustion, this evolutionary picture involves the star 
moving quickly across the HR diagram to the red supergiant region
where it undergoes core He burning. 
During the red supergiant phase the stellar envelope 
becomes fully convective, which would mix a certain amount of the 
core material into the envelope. The relative abundances of 
CNO in the red supergiant atmosphere would then change drastically. The
predicted final ratios of these elements will clearly depend on the mass of 
CN and ON-cycled core gas which is mixed into the envelope (assumed to
have the stars natal composition), and the mass of the envelope which 
is left at the beginning of the red supergiant phase. The envelope
mass will be affected by the amount of mass the star has lost during
its hot main-sequence 
phase through radiatively driven winds. This global picture is 
simply and clearly described in Lamers et al. (\cite{lam2001}). 
A single major mass-loss event is then invoked to explain the 
nebula and this dense, cool
gas should have a chemical composition similar to the polluted 
red-supergiant atmosphere. Hence the abundances in the ionized nebula around
Sher\,25 could be used to test the validity of this argument, however
there has been no accurate determination of abundances in the
nebula to date. The spectra presented by Brandner et al. (1997a and 1997b)
does not have enough wavelength coverage to estimate electron 
densities and temperatures consistently, hence no quantitative 
nebular abundances can be derived. These spectra do indicate that the 
[N\,{\sc ii}]/H$\alpha$ line strength ratio is somewhat higher than the 
background H\,{\sc ii} region, which Brandner et al. interpret as a 
sign of real nitrogen enhancement in the circumstellar gas, however 
a full and accurate nebular analysis is still required. After a short period
as a red supergiant, and after the mass-loss has occurred it has been
speculated that Sher\,25 (and other blue supergiants) return to the 
blue region of the HR diagram, undergoing what have been called 
``blue-loops''. 

\begin{figure}[h]
\epsfig{file=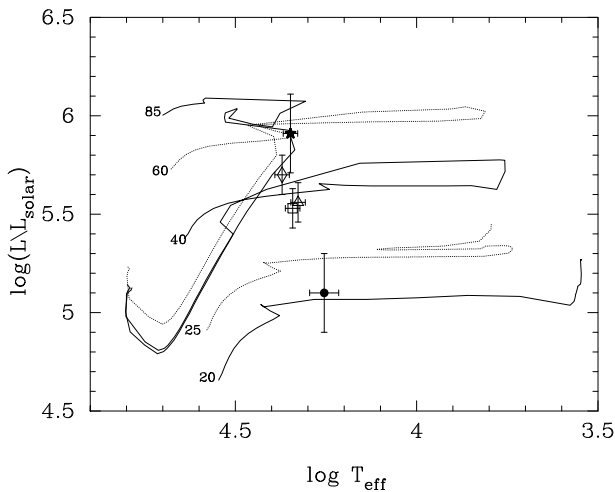,width=9cm}
\caption[]{Stellar evolutionary tracks for stars with zero-age main
sequence masses of 20, 25, 40, 60 and 85\msol from Meynet et al. 
\cite{meynet94}. The symbols are Sher\,25 (filled star), Sk$-$69$^{o}$202
progenitor of SN1987A (filled circle), HD2905 (open diamond), 
HD13854 (open square), HD14956 (open triangle). Luminosities
are taken from Paper\,II and  Walborn et al. (1989);  apart
from HD13854 which was calculated assuming membership of Per\,OB1,
distance as in Paper\,II, 
and optical magnitudes from the Simbad database. }
\label{evol_tracks}
\end{figure}

We can use the photospheric abundances we have derived for Sher\,25 
to test this model of a mass-loss event occurring when the star was
a red supergiant and subsequent bluewards evolution. Lamers et al.
({\cite{lam2001}) have recently presented comparisons between  
chemical abundances in LBV nebulae and what one would expect in
different evolutionary scenarios. They come to the conclusion that 
LBV nebulae are ejected during the blue supergiant phase, and
that such stars have {\em not} gone through a previous red supergiant 
episode. We follow their line of argument 
to determine if the CNO abundances we derive in Sher\,25 are consistent
with a previous red supergiant phase. The Si and Mg abundances
in Sher\,25 are very similar to those in the three solar neighbourhood
comparison stars, and these are not expected to be affected by any
mixing of H-burning products into the photosphere. The
Galactocentric distance of NGC3603 is approximately 8.5\,kpc, 
roughly the same as the Sun, and galactic abundance gradients
derived in Rolleston et al. (\cite{roll2000}) and 
Smartt \& Rolleston (\cite{smrol97}) indicate that there
is fairly low scatter in stellar metallicity at this distance from the 
Galactic centre. Hence we assume that Sher\,25 had an initial 
chemical composition similar to B-type stars and nebulae in the 
solar neighbourhood. Table\,\ref{CNO_comparison} lists the number
fractions of CNO derived in Sher\,25, the three comparison 
supergiants, the Sun, and solar neighbourhood B-type stars. 
From Table\,\ref{CNO_comparison} and Table\,\ref{mean_abun} it is
apparent that Sher\,25 has a nitrogen abundance approximately a 
factor of two lower than  
HD14956, and a factor of two higher than HD13854. These two 
stars have been classified in Paper\,I as 
``highly processed'' in that they are amongst the 
group of Galactic supergiants that show the highest N/C line strength 
ratios. The line strengths of Sher\,25 are comparable to stars
in this category, and the abundances we derive place it 
intermediate between HD14956 and HD13854 in terms of N enrichment. 
However it is certainly not extreme in its nitrogen abundance. 
It has a much higher abundance than HD2905, however we chose this
star as a comparison specifically because it was one of the objects in 
Paper\,I with the lowest apparent N enrichment. This 
illustrates the wide variation in nitrogen abundances that is 
seen amongst early B-type supergiants, and the 
abundance in Sher\,25 is similar to those Galactic stars which 
appear most N enhanced. 

Lamers et al. (2001) have presented the envelope abundances
that are predicted by the models of Meynet et al. (1994) 
after convective mixing in the red supergiant phase. This depends
critically on two factors. Firstly it depends on the mass-loss
during earlier phases, which is dominated by radiatively driven
winds while the star is hot. Clearly the higher the mass-loss, the
more of the envelope is lost and hence there is less gas of
initial composition to mix with the CNO-processed core material. 
Secondly it will depend on how far into the core the convection 
reaches and how much core gas actually gets dredged up into the 
convective atmosphere. Lamers et al. (2001) have plotted 
the N/O and N/C ratios for various values of mass-loss rates 
and initial stellar mass. We estimate the 
zero-age main-sequence mass of Sher\,25 to have been 60\msol\,
from comparison with the Meynet et al. tracks in Fig.\,\ref{evol_tracks}. 
Although we have directly derived a mass-loss rate for Sher\,25 in 
its present status as a luminous B-type supergiant, it may have 
had a higher $\dot{M}$ when it was a main sequence star. 
A 60\msol\ star on the main-sequence would have a mass-loss 
rate of typically $\sim2-5\times10^{-6}$\msol\,yr$^{-1}$ (Puls et al. 
\cite{puls96}), hence during its H-burning lifetime 
(4\,Myrs; from the Geneva models) 
it would have lost 8-20\msol\ of its envelope through its stellar wind. 
The surface N/O and N/C ratios show a steep dependence on 
$\dot{M}$ in this range, and the Lamers et al. calculations
indicate ranges of 
$3 \lesssim {\rm N/O} \lesssim  20$  and
$6 \lesssim {\rm N/C} \lesssim  30$. 

In order to compare this meaningfully with the abundance ratios we
derive, we must have confidence in the absolute abundances derived in 
our methods. To date the evaluation of this has been difficult, as 
pointed out in Smartt et al. (\cite{sma2001a}, \cite{sma97})
and Paper\,I. These studies have concentrated on 
differential analyses and comparisons of the difference between 
stellar subgroups, rather than assuming validity of the
absolute abundances. In particular the absolute value of the 
carbon abundance in our stars is of dubious validity, as 
we discussed in detail in Sect.\,\ref{photosphere}. Hence we will 
not quantitatively compare the N/C ratios in our stars to those in the 
evolutionary models. However as discussed in Sect.\,\ref{photosphere} 
the N and O absolute results should be reliable enough to 
carry out a meaningful comparison, providing we consider the 
probable errors in the methods. 
In Table\,\ref{CNO_comparison}
the N/O fraction is listed for each star. Given the uncertainties in 
the individual abundances, 
these ratios are uncertain within a factor of approximately 3. 
For Sher\,25, the observed N/O ratio (0.36) is much lower than that expected 
from the Lamers et al. (2001) calculation for a star which has undergone
convective dredge in the red supergiant phase (3$-$20), even allowing for 
an uncertainty of 3 in our derived results. For the most
nitrogen rich star in our sample HD14856 (of initial
mass approximately 40\msol) N/O is predicted to be greater
than 2, compared with our estimate of 0.53. This also implies the 
observed N/O ratio is too low to be compatible with convective
dredge-up having occurred. The global analysis of the 
large sample of B-type supergiants in Paper\,I implied that
their CNO line ratios and estimated abundances were not consistent
with them (or even a subset of them) being post red supergiant
stars. Our analysis of Sher\,25 indicates that it has similar
CNO abundances to the ``highly processed'' stars, but that it
is not extremely nitrogen rich. This suggests that, despite
the appearance of the nebula, the star has {\em not} undergone 
dredge-up and thus has not previously been a red supergiant. 

\subsection{Evolution direct from the main sequence}

Although we see no strong evidence for a previous red-supergiant
stage of Sher\,25, the abundances do suggest that some form of CN and 
NO cycle processed gas has been mixed into the stellar atmosphere. 
The sample in Paper\,I indicates that there is a range in the
CNO abundances in stellar atmospheres. This is consistent 
with rotationally induced mixing while the stars are 
main-sequence O-types and/or as they evolve off the main-sequence
to the cooler surface temperatures that we presently see. In this 
case one would envisage a range of C/N and N/O ratios in the 
atmospheres due to the distribution of rotational velocities. 
Lamers et al. (2001) have also estimated the C/N and N/O surface abundances
for massive stars that undergo rotationally induced mixing. 
They calculate the surface abundance ratios as a function of the 
mixing time $\tau_{\rm mix}$ and find that their results 
agree well with the calculations from
Heger \& Langer (\cite{heg2000}) and Maeder (\cite{mae98}) if 
$\tau_{\rm mix} \simeq1-3 \tau_{\rm MS}$ (where $\tau_{\rm MS}$ 
is the main-sequence lifetime). Further they estimate that  
$\tau_{\rm mix}$ = 2-4 $\tau_{\rm MS}$ is needed to account 
for their observed LBV nebular abundance ratios. 

For a 60\msol\ star the N/O ratio  in the atmosphere 
of a blue supergiant which has undergone rotationally induced mixing 
is in the range of $1<$N/O$<3$ if $\tau_{\rm mix}$ is $2-4$ times the main
sequence lifetime. The  value for Sher25 (N/O=0.36) is lower 
than these predicted values for  $2 < \tau_{\rm mix} < 4$, but is 
a factor of 3 higher than the expected initial value of N/O=0.12. Our
measured value suggests that Sher25 has undergone less mixing 
than these models predict, although almost certainly some moderate 
amount of mixing has occurred to enhance the N abundance. This 
suggests a larger
mixing time (i.e. greater than $2-4$) is more appropriate for Sher25. 

\section{Conclusions and wider implications}

\subsection{Implications for the evolutionary status of blue supergiants}

From our detailed abundance analysis of Sher\,25 and comparison to 
evolutionary calculations we can find no clear evidence that the 
star was previously in a red supergiant phase. The CNO abundances
we derive at its surface are similar to those in other 
Galactic B-type supergiants which show somewhat enhanced N/C 
and N/O abundance ratios. However the extent of this enhancement
does not appear to be high enough to suggest the star
dredged up CN and NO cycle equilibrium core gas during 
a cool supergiant phase. Even allowing for a large mass-loss rate 
during its main-sequence lifetime does not reconcile the results, 
and the current wind properties of Sher\,25 also indicate that it
is a relatively ``normal'' B-type supergiant. Certainly this
star sits comfortably in the largest sample of Galactic B-type supergiants
yet analysed (Paper\,I and Paper\,II), in terms of its mass, atmospheric parameters, 
abundances and wind properties. It does appear to be similar to the 
group of supergiants which show the largest N/C line strengths, 
and hence N/C abundance ratios but our quantitative comparison of the 
N/O results are more compatible with it having undergone some form
of rotationally induced mixing either during or just after its
main-sequence phase. 

The sample of Galactic A-type supergiants analysed by Venn (\cite{venn95})
shows evidence for the stars having undergone partial mixing 
of CN-cycled gas. However the abundances derived are not suggestive 
of the stars having been through a first dredge up and hence are
unlikely to be post red supergiant objects. Venn concludes that these
5-20\msol\ stars have evolved directly from the main-sequence. The 
Galactic B-type supergiants analysed in Paper\,I are of higher
mass ($\sim$20-40\msol), and the results are also not consistent
with them having undergone a previous red supergiant evolutionary
phase. The study of the abundances in the nebulae around LBVs
(in the 40-80\msol\ mass range) by Lamers et al. (2001) again 
suggests that they have not previously been red supergiants and 
that the nebulae are ejected during the blue supergiant phase. 
The chemical enhancements seen in these three cases are possibly
all due to some form of rotation induced mixing, which will have
some broad distribution based on initial rotational velocities 
and mass-loss rates. 

We find that the blue supergiant Sher\,25 is also very unlikely to 
have gone through a red supergiant phase. Compared to other 
B-type supergiants in the galaxy, its atmospheric chemistry and
wind properties are not peculiar in any way. Its peculiarity 
stems from the existence of the striking circumstellar nebula. 
Lamers et al. (2001) have speculated that the nebulae around 
LBVs have been ejected in the blue supergiant phase, although 
what the mechanism actually responsible for this is open to some debate. 
Therefore in the Galaxy, there is no single blue supergiant
or nebula surrounding a blue supergiant, which displays 
abundances consistent with the star having evolved through a 
previous red-supergiant phase. The idea that nebulae 
surrounding massive, evolved
stars must have been ejected during the red supergiant phase
does not appear to be supported by stellar analyses of blue 
supergiants shown here, in Paper\,I and in Lamers et al. (2001). 

\subsection{Comparison with SN1987A}
The origin of Sher\,25's nebulae is thus not well understood. 
Brandner et al. (1997b) have estimated the dynamical age of
both the equatorial ring and bi-polar lobes to be around 6600\,yr, 
and the total ionized gas to be between $0.3-0.7$\msol. 
The current mass-loss rate is much too low to have had much impact on 
the mass of this material, as in 6600\,yr the mass of gas ejected
in the current fast wind is $\sim$0.01\msol. 
Brandner et al. (\cite{bran97a}, \cite{bran97b}) have 
also  speculated that the Sher\,25 nebula is 
similar in morphology, mass, composition and kinematics to that
now seen around SN1987A. The SN1987A nebulae was almost certainly 
there before explosion, obviously too faint and compact to have been
detected on pre-explosion plates, suggesting the progenitor star
Sk$-$69$^{o}$202 was at a similar evolutionary state to Sher\,25
(although somewhat lower mass). By implication
our results then suggest that Sk$-$69$^{o}$202 
may never have gone 
through the red supergiant phase, but rather evolved directly from the
main-sequence and ejected its circumstellar material during the 
blue supergiant phase. 

The initial ejecta of SN1987A was studied by Fransson et al. 
(\cite{fransson89}) with IUE and ratios of N/C$\simeq$7.8
and N/O$\simeq$1.6 were found. The latter is significantly more than that
in the atmosphere of Sher\,25 and could be compatible with the
progenitor having had a previous red supergiant phase from comparisons
with the Lamers et al. calculations. However the modeling of this
emission line spectrum is problematic and uncertainties in N/O are
probably of the order $\pm$1. 
Three features of the SN1987A event have caused problems for
theoretical models looking for a consistent explanation of the
event; the relatively high temperature of \sk ; the stellar ring around
the SN (Wampler et al. \cite{wamp90}, and Jakobsen et al. \cite{jac91});
and the chemical anomalies discussed by Fransson et al. (\cite{fransson89}). 
The review by Podsiadlowski (\cite{pod92}) suggests that no current 
theory of the evolution of single stars can simultaneously cope with all
three and that binary accretion (or merger model) is theoretically
the most promising, especially with regard to the mechanism for ring 
formation. We find no evidence for radial velocity variations of Sher\,25 
(within the uncertainties of $\sim$7\,\kms) 
although the four 1200s exposures 
were taken back to back. We plan future velocity monitoring of this star 
to determine whether or not it is a binary system. In the wider picture
the question of how many blue supergiants are supernova
Type\,II progenitors is very uncertain although new projects 
with data mining and the rich data archives which now exist may shed 
some light on this in the future (e.g. Smartt et al. \cite{sma2001b}). 
It would be also worthwhile to determine the CNO abundance
ratios in other early B-type supergiants in the LMC in order to gain 
insight into the stars of similar metallicity at a similar evolutionary
state to \sk.

\begin{acknowledgements}
Spectroscopic data were obtained at the 
Anglo-Australian Observatory in Siding Spring, New South Wales 
and we are grateful to the AAO staff for their assistance.
We made use of the SIMBAD database maintained at CDS, Strasbourg.
S. Smartt thanks PPARC for support through an Advanced Research
Fellowship award. 
F. Rosales and N. Wright acknowledge PPARC support for the 
International Undergraduate Summer School held at the 
University of Cambridge during June and July 2001, during which time
this work was carried out. We specially thank Margaret Penston
for her efforts in organising this Summer school. We thank the two 
referees Antonella Nota and Henny Lamers for their helpful comments. 
\end{acknowledgements}

\end{document}